\documentclass[3p,times,procedia]{elsarticle}
\usepackage{nupha_ecrc}

\volume{00}

\firstpage{1}

\journalname{Nuclear Physics A}

\runauth{}

\jid{nupha}

\jnltitlelogo{Nuclear Physics A}

\usepackage{amssymb}

\usepackage[figuresright]{rotating}

\begin{document}

\begin{frontmatter}



\dochead{XXVIIth International Conference on Ultrarelativistic Nucleus-Nucleus Collisions\\ (Quark Matter 2018)}

\title{SMASH -- A new hadronic transport approach}


\author[1,2,3]{Hannah Petersen}
\ead{petersen@fias.uni-frankfurt.de}
\author[4]{Dmytro Oliinychenko}
\author[2,1]{Markus Mayer}
\author[2,1]{Jan Staudenmaier}
\author[2]{Sangwook Ryu}

\address[1]{Frankfurt Institute for Advanced Studies, Ruth-Moufang-Strasse 1, 60438 Frankfurt am Main, Germany}
\address[2]{Institute for Theoretical Physics, Goethe University, Max-von-Laue-Strasse 1, 60438 Frankfurt am Main, Germany}
\address[3]{GSI Helmholtzzentrum f\"ur Schwerionenforschung, Planckstrasse 1, 64291 Darmstadt, Germany}
\address[4]{Lawrence Berkeley National Laboratory, 1 Cyclotron Rd, Berkeley, CA 94720, US}

\begin{abstract}
Microscopic transport approaches are the tool to describe the non-equilibrium evolution in low energy collisions as well as in the late dilute stages of high-energy collisions. Here, a newly developed hadronic transport approach, SMASH (Simulating Many Accelerated Strongly-interacting Hadrons) is introduced. The overall bulk dynamics in low energy heavy ion collisions is shown including the excitation function of elliptic flow employing several equations of state. The implications of this new approach for dilepton production are discussed and preliminary results for afterburner calculations at the highest RHIC energy are presented and compared to previous UrQMD results. A detailed understanding of a hadron gas with vacuum properties is required to establish the baseline for the exploration of the transition to the quark-gluon plasma in heavy ion collisions at high net baryon densities.
\end{abstract}


\end{frontmatter}


\section{Introduction}
\label{intro}
Hadronic transport approaches have been successfully applied to describe the dynamical evolution of heavy ion reactions since many years (see for example \cite{Buss:2011mx, Nara:2016phs, Cassing:2009vt, Bass:1998ca,Bleicher:1999xi}). In particular, the hadronic non-equilibrium dynamics is crucial for the whole/partial evolution of the system at low/intermediate beam energies as well as for the late dilute stages at high RHIC and LHC energies. The motivation to establish a new approach is to provide a flexible, modular approach condensing the knowledge aquired with existing approaches as well as incorporating new experimental data for elementary cross-sections and branching ratios. The goal is to provide baseline calculations with hadronic vacuum properties to identify signals of the phase transition to the quark-gluon plasma. 

\section{Validation of SMASH}
\label{valid}
The approach that is presented here is called SMASH (Simulating Many Accelerated Strongly-interacting Hadrons) and incorporates the well-established mesons and baryons up to a mass of 2 GeV as degrees of freedom \cite{Weil:2016zrk}. Binary interactions are taking place via resonance excitation and decay or string excitation and fragmentation. A geometric collision criterion is employed. As a first test of the collision finding algorithm, an analytic solution of the Boltzmann equation within a Friedmann-Robertson-Walker metric \cite{Heinz:2015gka} has been nicely reproduced \cite{Tindall:2016try}.

\begin{figure} [h]
\centering
\includegraphics[width=0.19\textwidth]{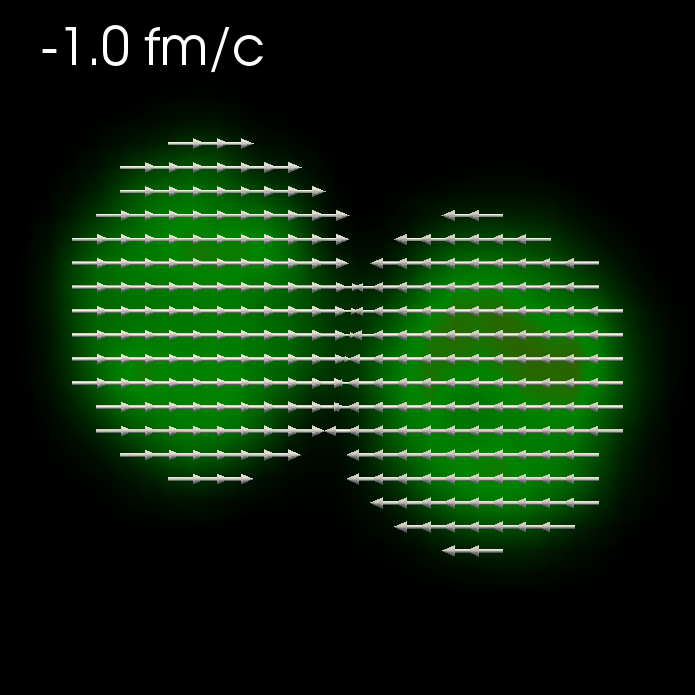}
\includegraphics[width=0.19\textwidth]{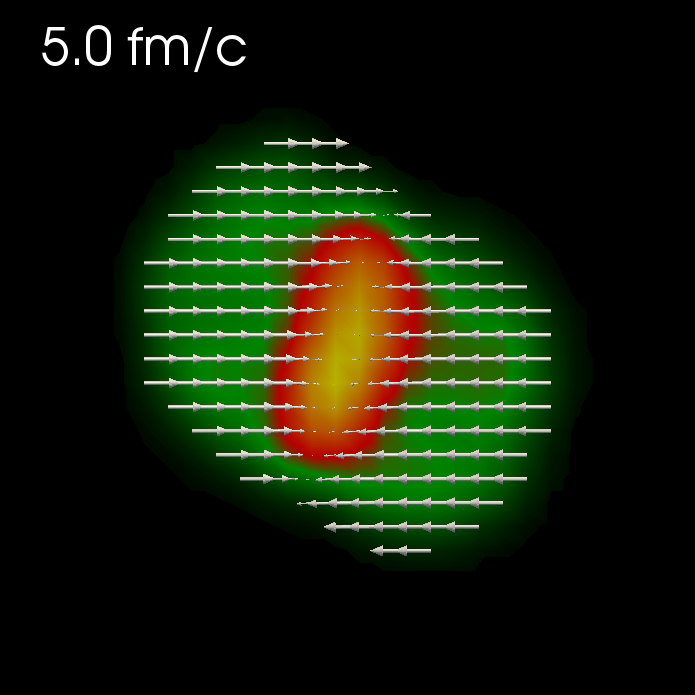}
\includegraphics[width=0.19\textwidth]{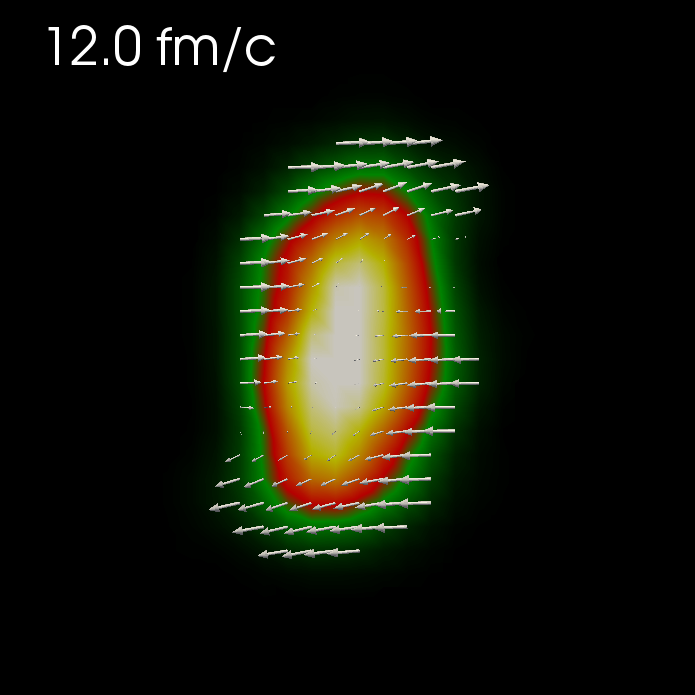}
\includegraphics[width=0.19\textwidth]{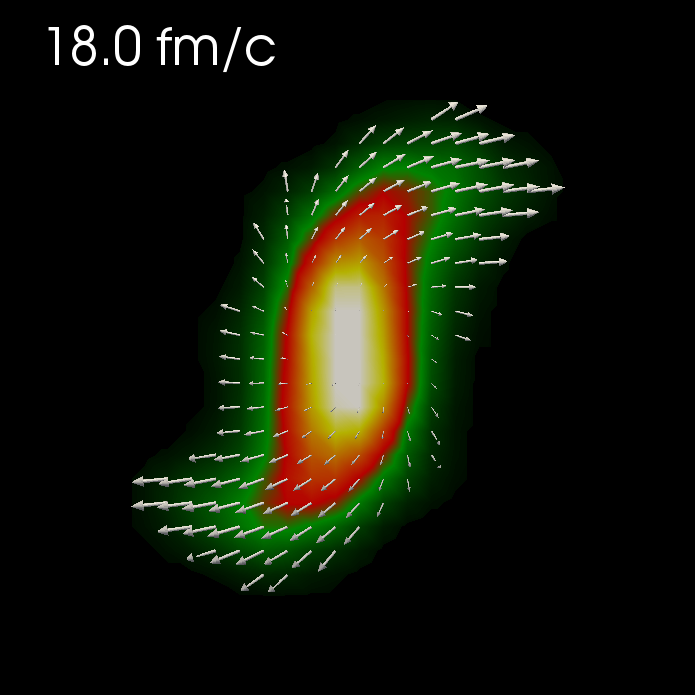}
\includegraphics[width=0.19\textwidth]{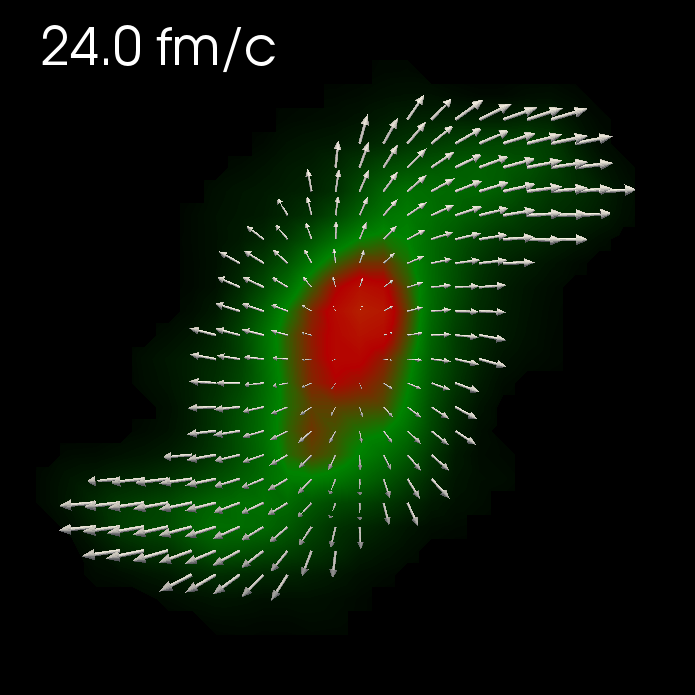}
\caption{
 Landau rest frame energy density $T^{00}_L$ (background color, up to $0.6$ GeV/fm$^3$) and velocity of Landau frame (arrows) in Au+Au collision at $E_{\rm kin} = 0.8A\,{\rm GeV}$ with impact parameter $b = 3$ fm, $N_{\rm test} = 20$. The velocity is proportional to the arrow length, the maximal arrow length corresponds to velocity of 0.55 $c$ (taken from \cite{Weil:2016zrk}).
}
\label{fig:landau_e_v}
\end{figure}

As a visual validation Fig. \ref{fig:landau_e_v} shows several snapshots of the time evolution of a Au+Au collision at $E_{\rm kin}=0.8A$ GeV at $b=3$ fm. The fireball of high energy density forms and a typical transverse flow profile is developed. The thermodynamic quantities are calculated from an event with 20 testparticles per real particle in the Landau frame. More quantitatively, cross-sections for elementary reactions are compared to experimental data as well as the check of detailed balance. In infinite matter calculations with various particle content, it has been shown, that the same amount of reactions takes place in forward and backward direction, even in each separate phase-space bin (see \cite{Weil:2016zrk} for examples).

\section{Bulk observables}
\label{bulk}
In \cite{Weil:2016zrk} it has been shown that the pion production in Au+Au collisions at $E_{\rm kin}=1-2A$ GeV describes the experimental data in a reasonable fashion. The inclusion of nuclear potentials slightly reduces the pion production whereas Fermi motion increases it and Pauli blocking again leads to a reduction of particle production due to the forbidden reactions. 

\begin{figure}[t]
\centering
\includegraphics[width=0.4\textwidth]{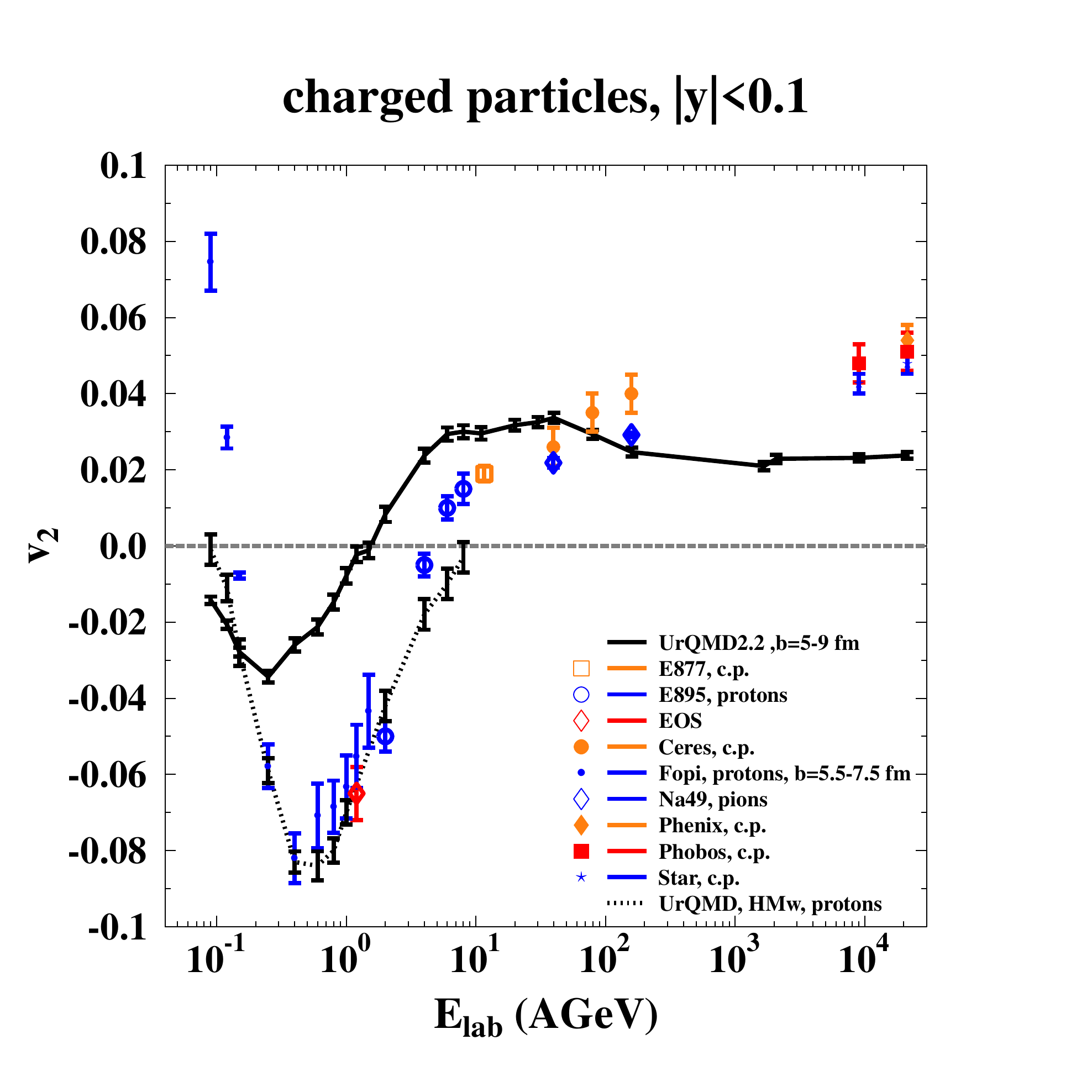}
\includegraphics[width=0.5\textwidth]{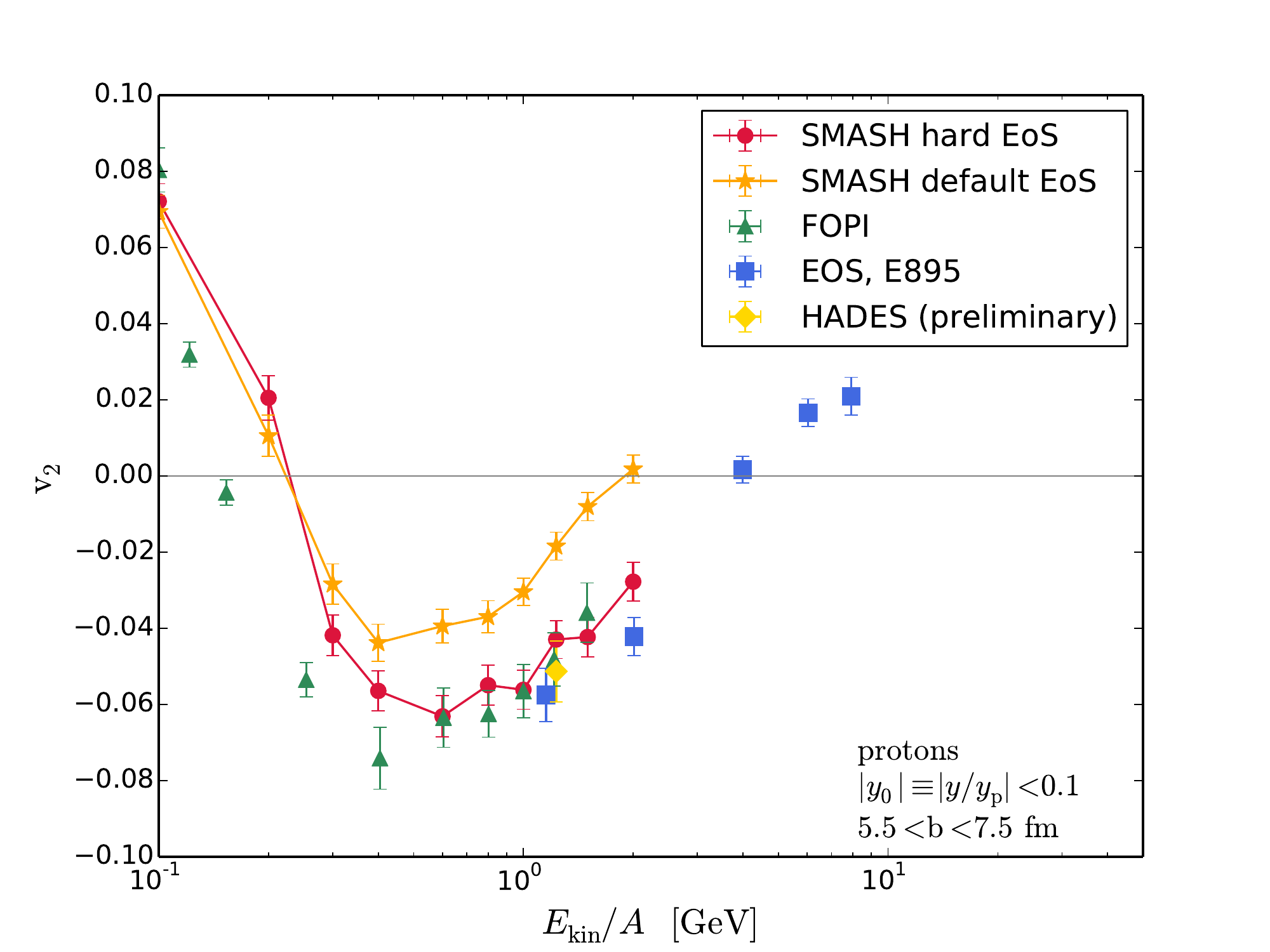}
\caption{Excitation function of elliptic flow compared to experimental data calculated within the UrQMD approach (left, Fig. taken from \cite{Petersen:2006vm}) and SMASH (right) employing different options for the equation of state. See \cite{Petersen:2006vm} for references to the experimental data.}
\label{fig:elliptic_flow}
\end{figure}

Fig. \ref{fig:elliptic_flow} shows the excitation function of elliptic flow over a large energy range. On the left, a calculation within the Ultra-relativistic Quantum Molecular Dynamics (UrQMD) approach \cite{Bass:1998ca,Bleicher:1999xi} is shown with respect to the experimental data. The full line indicates the result from the cascade approach, while the dashed line depicts the calculation including a mean field for the nucleons as described in \cite{Li:2006ez}. On the right hand side, the corresponding calculation within SMASH is shown including a Skyrme mean field corresponding to two different compressibilities. The default values are chosen according to \cite{Xu:2016lue}. Both calculations agree qualitatively that a harder equation of state reproduces the flow at low energies. 

In addition, the shear viscosity over entropy ratio has been calculated as a function of temperature and net baryon chemical potential and agrees well with the previous UrQMD result \cite{Rose:2017bjz}. In \cite{Oliinychenko:2016vkg} forced canonical thermalization in certain phase-space regions as a proxy for multi-particle collisions has been explored. It has been shown that SMASH with thermal bubbles interpolates between a pure hydrodynamic and a pure transport calculation.

\section{Electromagnetic probes and hadronic rescattering}
\label{em_resc}
\begin{figure}[b]
\centering
\includegraphics[width=0.48\textwidth]{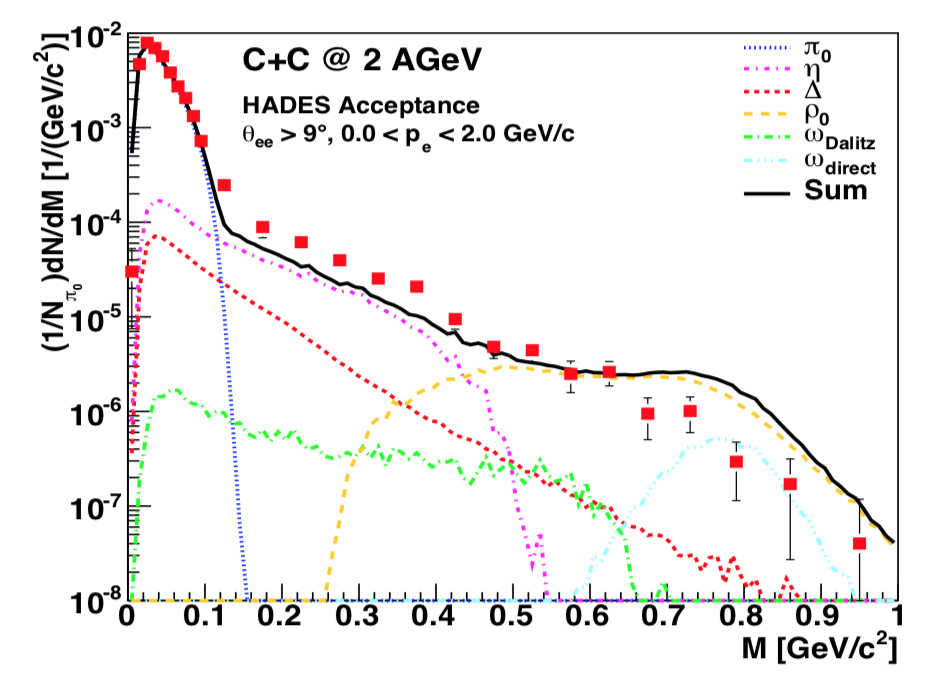}
\includegraphics[width=0.46\textwidth]{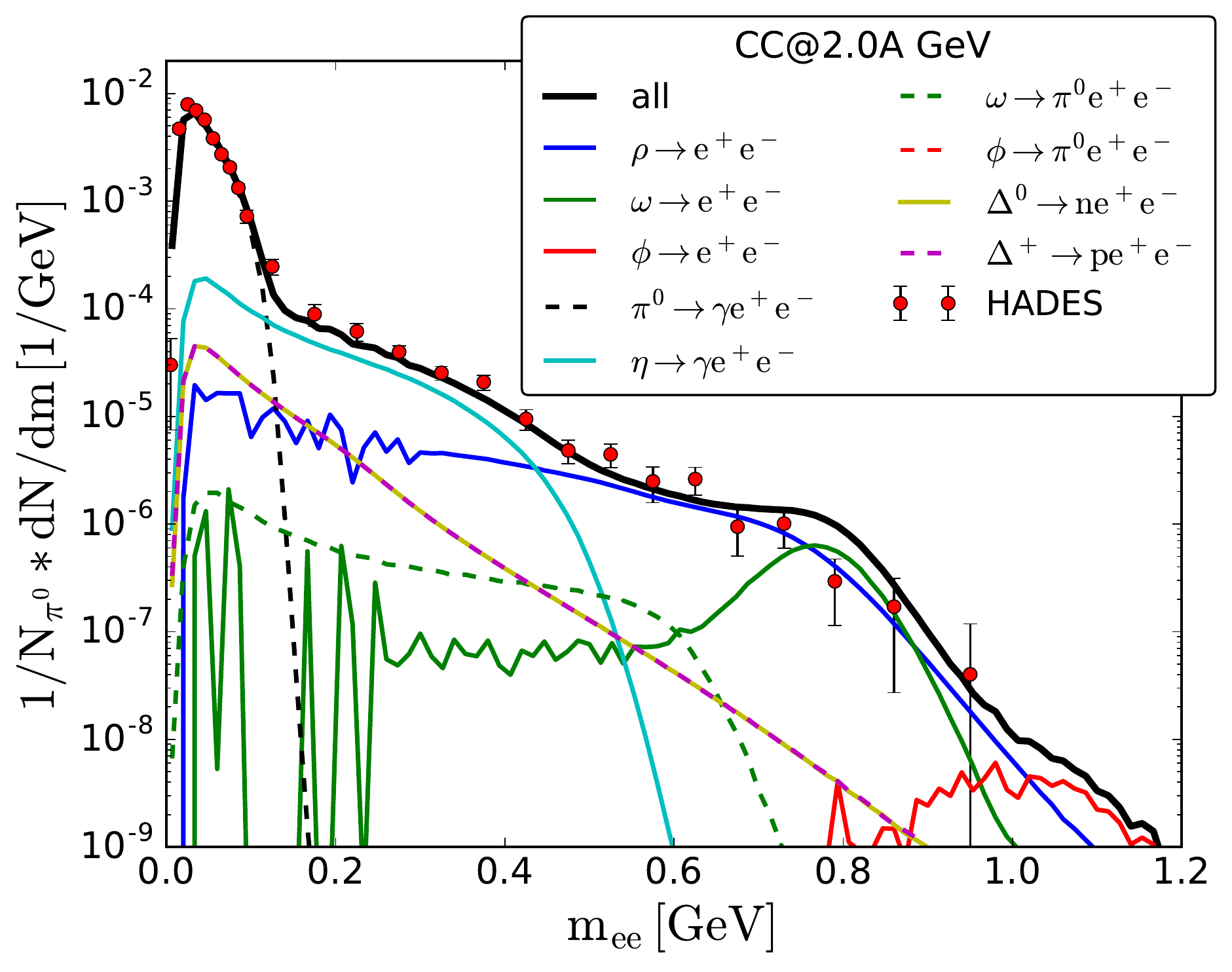}
\caption{Dilepton production as a function of the invariant mass in carbon-carbon collisions at $E_{\rm lab}=2A$ GeV within the UrQMD transport approach (left, Fig. taken from \cite{Endres:2013nfa}) and SMASH (right, Fig. taken from \cite{Staudenmaier:2017vtq}).}
\label{fig:dileptons}
\end{figure}

Complementary to the study of bulk observables, the whole set of dilepton measurements provided by the HADES collaboration has been explored in \cite{Staudenmaier:2017vtq}. In general, the hadron-resonance approach with vacuum properties provides a good description of the dilepton emission in elementary and small systems while in collisions of heavier ions the in-medium modifications of the spectral functions \cite{Rapp:1999us} are important. Fig. \ref{fig:dileptons} shows a side by side comparison of UrQMD calculations (left) and SMASH calculations (right). Overall, both approaches yield very similar results as expected. In SMASH, the branching ratios contributing to the $\rho$ meson peak have been adjusted and dilepton emission by the vector mesons below the hadronic threshold has been included, which improves the agreement in the low mass region. 

\begin{figure}
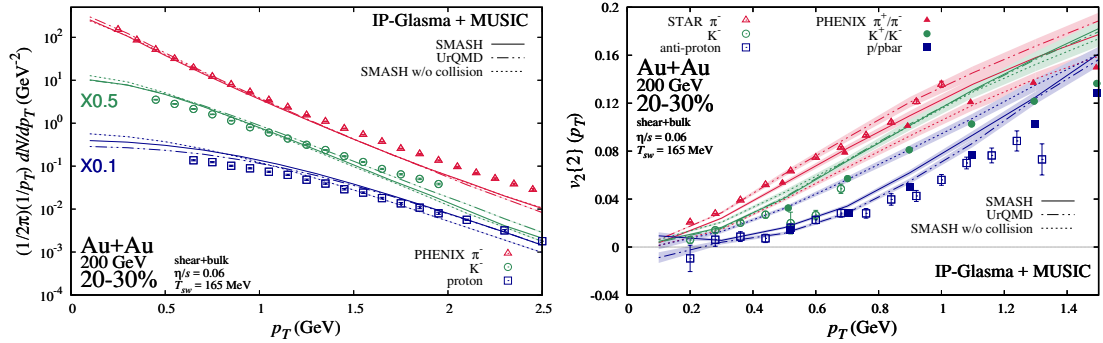

\centering
\includegraphics[width=0.48\textwidth]{dNdpT_spc_WMSample_nBBa_2030.pdf}
\includegraphics[width=0.48\textwidth]{v2pT_spc_WMSample_nBBa_2030.pdf}
\caption{Influence of hadronic rescattering on transverse momentum spectra and elliptic flow for identified particles in Au+Au collisions at $\sqrt{s_{\rm NN}}=200$ GeV within UrQMD \cite{Ryu:2017qzn} and SMASH.}
\label{fig:afterburner}
\end{figure}

Last but not least, Fig. \ref{fig:afterburner} shows a comparison of the hadronic rescattering within a hybrid framework at the highest RHIC energy. The protons receive an increase of transverse momentum and the mass splitting of elliptic flow is increased significantly. SMASH does not reach the same magnitude of the effects as within UrQMD due to missing additional baryon-antibaryon annihilation processes and cross-sections provided by the additive quark model for exotic combinations of hadrons. 

\section{Conclusions}
\label{concl}
Overall, it has been shown, that SMASH is a new hadronic transport approach, that describes bulk observables and dilepton emission at low beam energies and can be employed for the late stage hadronic rescattering at high energies. The results are comparable to the ones from the established similar UrQMD transport approach. 




\section{Acknowledgements}
H.P. acknowledges funding of a Helmholtz Young Investigator Group VH-NG-822 from
the Helmholtz Association and GSI. D. O. was supported by the U.S. Department of Energy, 
Office of Science, Office of Nuclear Physics, under contract number DE-AC02-05CH11231 and received support within the framework of the Beam Energy Scan Theory (BEST) Topical Collaboration. This work was supported by HIC for FAIR within the framework of the LOEWE program launched by the State of Hesse. Computational resources have been provided by the Center for Scientific Computing (CSC) at the Goethe-University of Frankfurt and the GSI GreenCube. This work was supported by the DFG through the grant CRC-TR 211.





\end{document}